\documentclass{article}

\PassOptionsToPackage{numbers, compress}{natbib}

    \usepackage[final]{neurips_2022_ml4ps}

\usepackage[utf8]{inputenc} 
\usepackage[T1]{fontenc}    
\usepackage{hyperref}       
\hypersetup{
    colorlinks,
    linkcolor={red!50!black},
    citecolor={blue!50!black},
    urlcolor={blue!80!black}
}
\usepackage{url}            
\usepackage{booktabs}       
\usepackage{amsfonts}       
\usepackage{nicefrac}       
\usepackage{microtype}      
\usepackage{xcolor}         
\usepackage{amsmath}
\usepackage{amssymb}
\usepackage{graphicx}
\usepackage{caption}
\usepackage{subcaption}

\newcommand{\prospector}{nested sampling}

\title{Monte Carlo Techniques for Addressing Large Errors and Missing Data in Simulation-based Inference}

\author{%
  Bingjie Wang
  \thanks{Secondary affiliations: Institute for Computational \& Data Sciences, and Institute for Gravitation and the Cosmos, The Pennsylvania State University, University Park, PA 16802, USA}\\
  Department of Astronomy \& Astrophysics\\
  The Pennsylvania State University\\
  University Park, PA 16802, USA \\
  \texttt{bwang@psu.edu} \\
  \And
  Joel Leja $^*$\\
  Department of Astronomy \& Astrophysics\\
  The Pennsylvania State University\\
  University Park, PA 16802, USA \\
  \texttt{joel.leja@psu.edu} \\
  \And
  Ashley Villar $^*$\\
  Department of Astronomy \& Astrophysics\\
  The Pennsylvania State University\\
  University Park, PA 16802, USA \\
  \texttt{vav5084@psu.edu} \\
  \And
  Joshua S. Speagle
  \thanks{Secondary affiliations: Dunlap Institute for Astronomy \& Astrophysics, and Department of Statistical Sciences, University of Toronto, Toronto, ON, M5S Canada}\\
  Department of Astronomy \& Astrophysics\\
  University of Toronto\\
  Toronto ON M5S 3H4, Canada\\
  \texttt{j.speagle@utoronto.ca}\\
}

\begin{document}

\maketitle

\begin{abstract}

Upcoming astronomical surveys will observe billions of galaxies across cosmic time, providing a unique opportunity to map the many pathways of galaxy assembly to an incredibly high resolution. However, the huge amount of data also poses an immediate computational challenge: current tools for inferring parameters from the light of galaxies take $\gtrsim 10$ hours per fit. This is prohibitively expensive.
Simulation-based Inference (SBI) is a promising solution. However, it requires simulated data with identical characteristics to the observed data, whereas real astronomical surveys are often highly heterogeneous, with missing observations and variable uncertainties determined by sky and telescope conditions. Here we present a Monte Carlo technique for treating out-of-distribution measurement errors and missing data using standard SBI tools.
We show that out-of-distribution measurement errors can be approximated by using standard SBI evaluations, and that missing data can be marginalized over using SBI evaluations over nearby data realizations in the training set.
While these techniques slow the inference process from $\sim 1$ sec to $\sim 1.5$ min per object, this is still significantly faster than standard approaches while also dramatically expanding the applicability of SBI. This expanded regime has broad implications for future applications to astronomical surveys.
\end{abstract}

\section{Introduction}

Advancement in the understanding of galaxy formation and evolution comes from two frontiers: increasingly large and/or advanced astronomical surveys, and increasingly sophisticated models to infer physical properties from observations. In the near future, surveys conducted by the Vera C. Rubin Observatory, among others, will increase our total inventory of surveyed galaxies across cosmic time from millions to billions. On the other hand, the current state-of-the-art tools for parameter inference typically involve Bayesian inference using a Markov chain Monte Carlo or nested sampling, which are prohibitively expensive for analyzing large data sets. This is because galaxies are inherently sophisticated systems, requiring the generation of $\sim$1-2 million models for each object to map out the complex likelihood surface. Generating one model typically takes $\sim$0.05s, translating to an expensive $\sim 100$ billion CPU-hours to fit the galaxies expected to be observed by the Rubin Observatory.

Simulation-based inference (SBI) is a promising solution to the computational challenge put forth by next-generation astronomical surveys. It bypasses a traditional likelihood framework to learn densities directly (see \cite{Cranmer2020} for a recent review). SBI-based methods have already begun to be adopted in the astrophysical literature (e.g., \cite{Alsing2019,Green2020,Zhang2021,Leja2022}). Moreover, SBI has been shown to be able to accurately approximate galaxy posteriors in a proof-of-the-concept study \cite{Hahn2022}. However, SBI has notable drawbacks: current methods require well-modeled noise properties and a complete set of input data, two assumptions which are often violated in real astronomical data.
First, uncertainties can fluctuate wildly due to varying telescope conditions, turbulent atmosphere perturbing photon paths, and/or the light from the galaxy of interest being contaminated by the light from its neighbors or foreground. Second, heterogeneous data coverage is common because data needs to be combined from multiple surveys, whose different telescopes and instruments do not perfectly overlap on the sky.

This paper presents a complete SBI-based methodology to handle noise outside of the training set and missing data. Specifically, we train a model using Amortized Neural Posterior Estimation to map the galaxy physical parameters onto photometry, employing normalizing flows as density estimators. Using this approach, we show that we can successfully reconstruct posteriors in the presence of large noise and missing photometric bands by marginalizing over available data in the training set.

\begin{figure}
  \centering
  \includegraphics[width=1\textwidth]{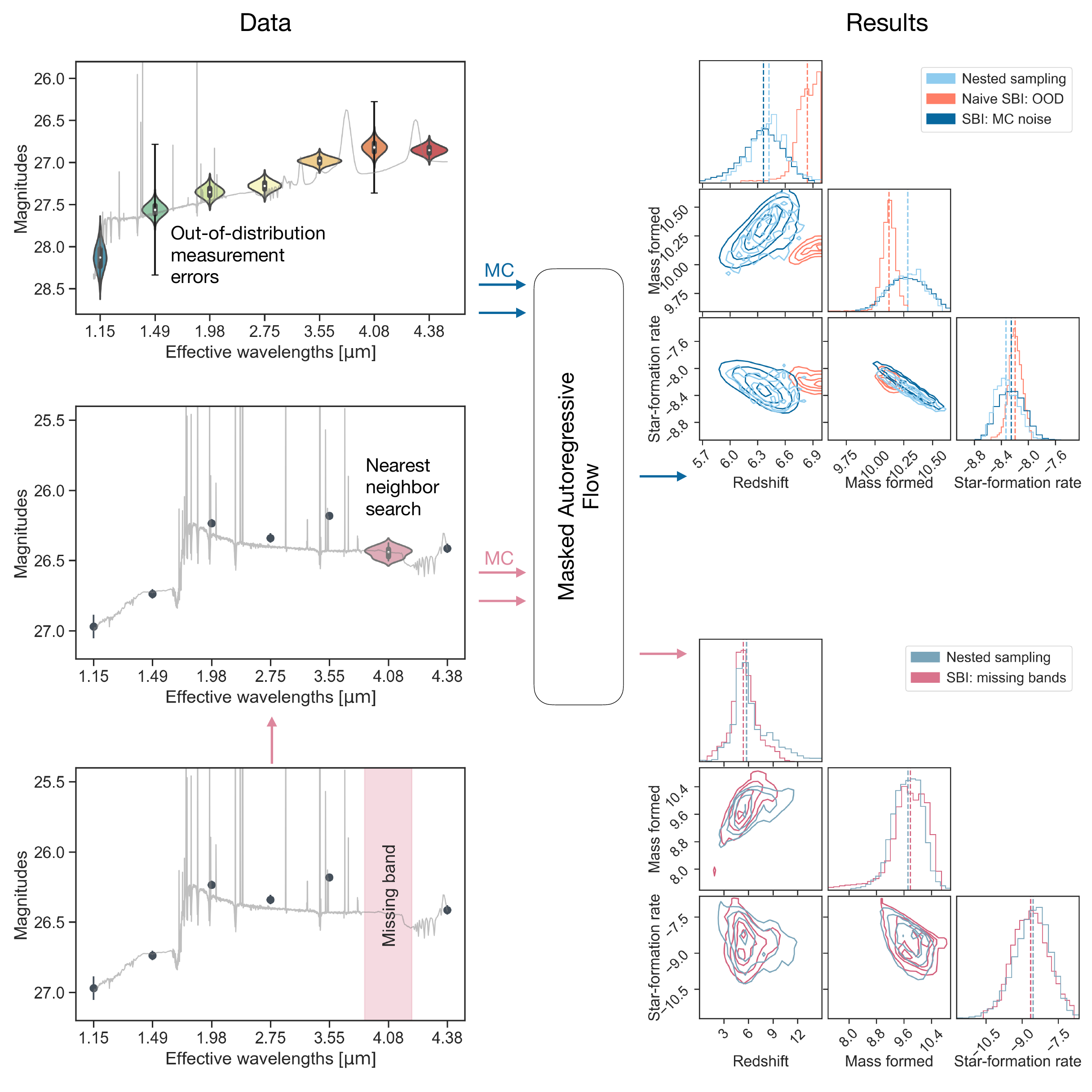}
  \caption{Schematic diagram showing our procedure for dealing with out-of-distribution (OOD) measurement errors and missing data. First, the violin plot on the top left shows one of our simulated SEDs, with Gaussian noise added to the true underlying SED. Given OOD uncertainties (black error bars), we marginalize over possible noise by Monte Carlo (MC) integration (Section~\ref{sec:mcnoise}). The top right corner plot shows the different posteriors from nested sampling, the naive usage of baseline SBI, and SBI with MC noise using the method presented here. Notably, our method performs similarly to the traditional method of nested sampling and markedly better than the naive SBI. Second, the SED in the bottom left panel has one band missing, rendering it inaccessible to our baseline SBI. Its approximate solution (middle left panel) is found by nearest neighbor search along with MC integration (Section~\ref{sec:missdata}). The resulting posteriors (bottom right) show good agreement with nested sampling. \label{fig:sche}}
\end{figure}

\section{Experiments\label{sec:exp}}

The training set consists of $\sim$2 million sets of model spectral energy distributions (SEDs) and the corresponding galaxy properties including distance, mass, age, gas composition, and star formation history. The validation set contains around 200 held-out examples, drawn from the same distribution as the training set. The size of the validation set is limited by the computational time required to estimate posterior quantities with nested sampling, which takes $\gtrsim 10$ hours per fit.  We simulate mock photometry for 7 bandpasses on the James Webb Space Telescope (JWST)
using a delayed--$\tau$ model as implemented in \texttt{Prospector} \cite{Johnson2021}. It consists of 7 free parameters describing the contribution of stars, gas and dust \cite{Ciesla2017,Carnall2019,Chevallard2019}. The surveyed parameter space roughly follows a mock catalog designed for JWST surveys \cite{Williams2018}. 
The noise is propagated into the training set by assuming a Gaussian noise distribution in asinh-magnitude-space.

We adopt the Masked Autoregressive Flow \cite{Papamakarios2017} implementation in the \verb|sbi| Python package\footnote{\url{https://github.com/mackelab/sbi/}}\cite{Greenberg2019,tejero-cantero2020}. The model has 15 blocks, each with 2 hidden layers and 500 hidden units. 
Training our model takes roughly a day on a single NVIDIA Tesla K80 GPU.

\section{Method\label{sec:method}}

In this section we detail the focus of this paper: the complete methodology to deal with out-of-distribution measurement errors and missing bands. A schematic representation of the methods is shown as Figure~\ref{fig:sche}. Their derivation can be found in the Appendix.

\subsection{Out-of-distribution measurement errors\label{sec:mcnoise}}

While we expect most observational noise can be captured by a carefully chosen noise model, in practice even generous training sets will not cover the wide range of observed noise distributions.
To solve this problem, we propose to use baseline SBI to marginalize over possible noise values via simple Monte Carlo (MC) integration. 
First, after identifying an out-of-distribution measurement, we create a set of 100 simulated photometry drawing from a Gaussian distribution with a mean of the observed value and a standard deviation of the observed uncertainty. In principle, the number of samples required is subject to the complexity of the posterior distribution. Here we find 100 draws is sufficient for our purpose. Each simulated photometry is then assigned an uncertainty of the mean in the noise model at its magnitude. These measurements are passed through the baseline SBI model to produce posteriors; subsequently averaging over all the ``noisy'' posterior samples provides the final parameter estimations. 

\subsection{Missing data\label{sec:missdata}}

Here we describe a method to approximate missing data by using a nearest neighbor approximation in the training set.
First, we find all SEDs in the training set whose reduced-$\chi^2$ ($\chi^2_{\rm red} = \chi^2/(n_{\rm bands}-1)$) calculated with respect to the observed SED are less than or equal to 5. 
Second, we construct a kernel density estimation (KDE) from those nearest neighbors, weighted by the inverse of their Euclidean distances to the observed SED, for each of the missing bands.
Finally, we draw random samples from the KDE and pass them to the baseline SBI, and average over the posteriors.

A caveat to this approach is that we only marginalize over values which are included in our training set, effectively producing additional dependence on the accuracy of the model priors. The effect of Bayesian priors on parameter inference is a well-known challenge and not discussed further here.

\section{Results\label{sec:res}}

\subsection{Computational efficiency}

Baseline SBI, i.e., when the data naturally fall within the simulated training set, takes about 1 second per fit. This can be compared with traditional inference methods, e.g., generating models on-the-fly and performing nested sampling, which take $\gtrsim 10$ hours per fit. Not only is this already a $>10^4$ speed increase, but SBI also shows remarkably comparable performance to the traditional methodology. The proposed algorithms to extend SBI to cover out-of-distribution noise and missing data take $\sim 1.5$ minutes per fit due to the multiple draws required; while slower than baseline SBI, our method is still $\sim400\times$ faster than traditional methods.
The change in the runtime as a function of the number of noisy/missing bands will be addressed in a forthcoming paper, as it is expected to be dependent on multiple factors, such as the complexity of the posterior distributions.

\subsection{Assessing the accuracy of out-of-distribution noise approximation\label{sec:res_mcnoise}}

\begin{figure}
  \centering
  \includegraphics[width=1\textwidth]{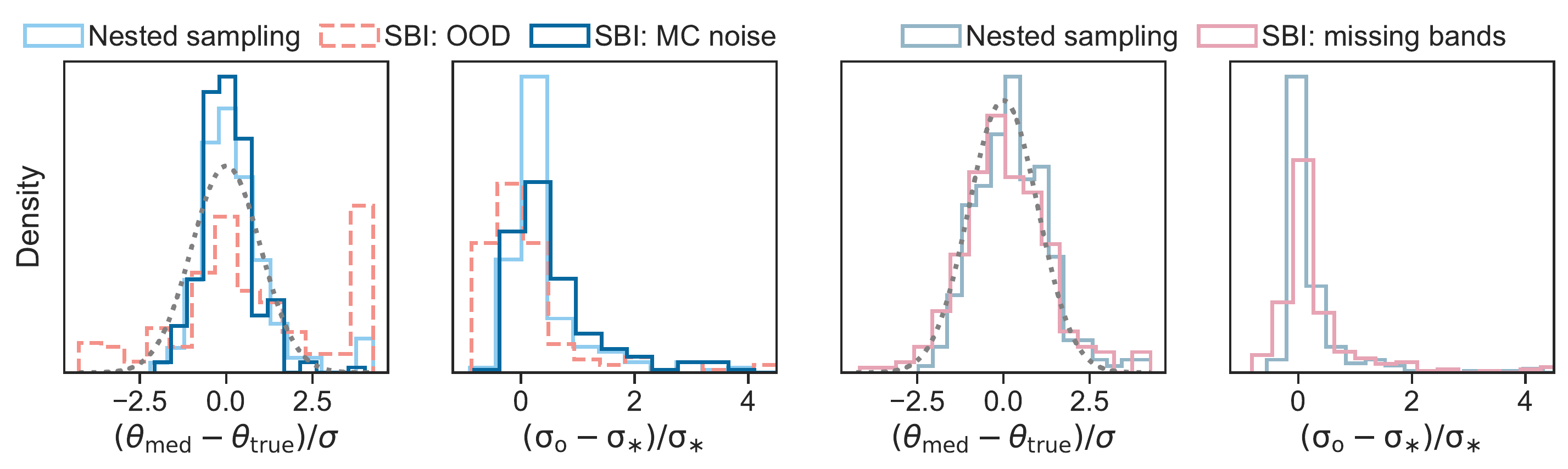}
  \caption{The two panels on the left illustrate the changes in SBI/\prospector\ posteriors estimated from the noisy photometry with respect to those from the unperturbed photometry for one of the parameters (redshift). 
  Similarly, the two panels on the right describe the changes in SBI/\prospector\ posteriors estimated from incomplete photometric data with respect to those from the complete data.
  Unit Gaussians are overplotted as gray dotted lines to guide the eye.
  It is evident that the methodologies proposed here, denoted by ``MC noise'' and ``missing bands,'' recover the parameters with accuracy comparable to standard inference methodology like \prospector. 
  We also show results from improperly using the baseline SBI when the noise is out-of-distribution (OOD) to demonstrate the necessity of applying our method. The $\delta_{\sigma,{\rm OOD}}<0$ group manifested in the second orange histogram shows naive SBI finds the wrong solution but with high confidence.
  \label{fig:hist}}
\end{figure}

In order to assess the accuracy in parameter recovery, we inflate the noise by $5 \sigma$ in two random bands for 200 test objects, and compare the shifts in medians and standard deviations of the posteriors generated from SBI (baseline) and SBI (MC noise). We also compare to the shifts in posteriors from \prospector\footnote{\url{https://github.com/joshspeagle/dynesty}} \cite{Speagle2020,sergey_koposov_2022}. 
It is well-known that statistical tests in multivariate settings is difficult. We thus choose this simple approach to evaluate whether the shifts seen in the SBI are expected over other tests such as KL-divergence.
The shift in medians is quantified as $\delta_{\rm med} = (\theta_{\rm med} - \theta_{\rm true}) / \sigma,$ where $\theta$ is the parameter of interest, and $\sigma$ is the $(84^{\rm th} - 16^{\rm th})/2$ quantile width in the posterior distribution. The shift in standard deviations is estimated as $\delta_{\sigma} = (\sigma_{\rm o} - \sigma_{\ast}) / \sigma_{\ast},$ where $\sigma_{\rm o}$ is the standard deviation of posteriors predicted from the noisy photometry, and $\sigma_{\ast}$ is that from the original photometry. 
The results for one of the parameters, redshift, are shown in Figure~\ref{fig:hist}. Other parameters exhibit similar trends. It is evident that the parameter recovery by our proposed technique, SBI (MC noise), is comparable to that of \prospector, while naively passing the out-of-distribution errors through the baseline SBI performs substantially worse as expected. 

We note that the MC process may generate noisy data which lie outside of one's model space entirely, causing dangerous extrapolation within the SBI machinery. To avoid this problem, we truncate a given Gaussian noise distribution to be within a range that is determined based on nearest neighbors chosen in the same way as in Section~\ref{sec:missdata} in magnitude space. In the occasional case where there is an insufficient number of neighbors ($n\leq10$) satisfying $\chi^2_{\rm{red}}\leq5$, we increase the cut on $\chi^2_{\rm{red}}$ in increments of 2.

\subsection{Assessing the accuracy of the nearest neighbor search for missing bands\label{sec:res_missdata}}

We similarly assess the accuracy of our missing data methodology by randomly masking a band for 200 test objects, and comparing the shift in medians and standard deviations of the posteriors. The results are also shown in Figure~\ref{fig:hist}. The fact that SBI (missing bands) and \prospector\ produce similar distributions in these spaces validates this approach. However, there are a number of cases where $\delta_{\sigma}$(SBI) is slightly greater than that of $\delta_{\sigma}$(\prospector). Upon close examination, we find that this occurs in multi-modal posteriors. While in most cases both solutions are captured, in some cases the nearest-neighbor approximation favors a particular mode. This behavior has two sometimes-overlapping causes. First, the training set can by chance be sparsely sampled in the parameter space where the secondary solution is, and hence it is difficult to find nearest neighbors that can produce this solution. This can be solved by more densely sampling parameter space in the training set. Second, our priors can explicitly disfavor the secondary solution, meaning few or no models exist there. This is a generic problem in SBI, as the training set must be generated following the prior density; the fact that the training set is also used to approximate missing bands makes the Bayesian priors doubly important in this technique.

\section*{Broader impact}

The SBI method presented here will be more broadly applicable to a wide range of fields---particularly those which also suffer from missing data or rapidly changing noise properties. Furthermore, SBI provides a ``greener'' solution to traditional inference problems, requiring notably less energy (CPU/GPU hours) to effectively reproduce a well-calibrated posterior. As noted in the paper, a word of warning is provided to the reader: naive applications of our proposed method of including MC noise may allow ones to go beyond the noise properties explored in the training set. In those cases where the data extends beyond the model set itself, the results will very likely be poorly calibrated.

\begin{ack}

B.W. is supported by the Institute for Gravitation and the Cosmos through the Eberly College of Science. This research received funding from the Pennsylvania State University’s Institute for Computational and Data Sciences through the ICDS Seed Grant Program. 
Computations for this research were performed on the Pennsylvania State University's Institute for Computational and Data Sciences' Roar supercomputer.

\end{ack}

\bibliographystyle{IEEEtranN}
\bibliography{sbi_neurips_wang.bib}


\appendix

\section{Appendix}

We present the mathematical framework of our proposed methods here. To start, we note that SBI bypasses a traditional likelihood framework to learn densities directly. This means we need access to a simulator function $\mathcal{S}_x(\theta)$ that can take in some input parameters $\theta$ and then generate some output data $x$; i.e., that we can generate independent and identically distributed (iid) such that
\begin{equation}
    \{x_{i,1},\dots,x_{i,j},\dots\} \overset{\textrm{iid}}{\sim} \mathcal{S}_x(\theta_i),
\end{equation}
where $\theta_i$ is a particular parameter, and $x_{i,j}$ is a particular realization of the data from the parameter. There is no guarantee that $\mathcal{S}_x(\theta)$ is analytic or even deterministic; in other words, it may not be possible to write down a likelihood $P(\theta_i|x_{i,j})$. However, if we have a large dataset of $n$ parameter-data pairs $\{\theta_i, x_i\}_{i=1}^{n}$, then we could consider using some machine learning method with hyperparameters $\phi$ to try and just learn the joint density directly:
\begin{equation}
    \{\theta_i, x_i\} \hookrightarrow P_{\phi}(\theta, x) \approx P(\theta, x),
\end{equation}
where we have explicitly included the $P_\phi(\cdot)$ notation to emphasize that this is an approximation to the true density $P(\cdot)$. This can be converted to a likelihood using Bayes' Theorem as
\begin{equation}
    P(x|\theta) \approx P_\phi(x|\theta) = \frac{P_\phi(\theta, x)}{P(\theta)},
\end{equation}
assuming $P(\theta)$ is known and/or can be approximated via $P_\phi(\theta)$. We can likewise derive the posterior under the same assumptions for $P(x) \approx P_\phi(x)$ via
\begin{equation}
    P(\theta|x) \approx P_\phi(\theta|x) = \frac{P_\phi(\theta, x)}{P_\phi(x)} \propto P_\phi(\theta, x),
\end{equation}
since $P_\phi(x)$ will be a constant for any individual object with data $x_i$.

Measurement errors imply that the data we observe, $x_i$, are actually different from the true data, $x_i^*$. Let us assume that for each point we can say without loss of generality that the probability distribution function (PDF) depends on some known measurement error, $\sigma_i$, such that the noisy measurement can be modeled via
\begin{equation}
    x_i \sim P(x_i|x_i^*, \sigma_i).
\end{equation}
The corresponding posterior is now equivalent to
\begin{equation}
    P(\theta|x,\sigma) = \int_{\Omega(x^*)} P(\theta|x^*) P(x^*|x,\sigma) \, {\rm d} x^*,
\end{equation}
where $\Omega(x^*)$ signifies the domain of $x^*$.

SBI can deal with noise contained inside of the training set. This is done by injecting the errors into the training set, and then conditioning on them. In other words, our simulator just becomes a function of both the input parameter $\theta$ and the measurement uncertainties $\sigma$ such that
\begin{equation}
    \{x_{i,1},\dots,x_{i,j},\dots\} \overset{\textrm{iid}}{\sim} \mathcal{S}_x(\theta_i,\sigma_i).
\end{equation}
This allows us to generate $n$ $\{\theta_i, x_i, \sigma_i\}_{i=1}^{n}$ pairs, which are then used to learn the joint density using the same strategy as above via
\begin{equation}
    P(\theta|x,\sigma) \approx P_\phi(\theta|x,\sigma) = \frac{P_\phi(\theta, x,\sigma)}{P_\phi(x,\sigma)} \propto P_\phi(\theta, x, \sigma).
\end{equation}

However, when we want to fit an object with out-of-distribution measurement errors, or even worse, with missing data, then it is not feasible. Below we describe how to deal with these situations.

If the measurement properties, $\sigma_i$, are outside of those that can feasibly be modeled, then we need to evaluate the integral over $x^*$. Using Bayes' Theorem and refactoring a few terms, this means we need to solve
\begin{equation}
    P_\phi(\theta|x,\sigma) \propto \int_{\Omega(x^*)} P_\phi(\theta, x^*) P(x | x^*, \sigma) \, {\rm d} x^*,
\end{equation}
where $P(x)$ is a constant that can often be ignored, $P_\phi(\theta, x*)$ is the PDF derived from $\{\theta_i, x_i^*\}$ pairs, and $P(x | x^*, \sigma)$ is the possibly unknown and/or analytically intractable PDF associated with the noise process.

Considering a general case where $P(x^* | x, \sigma)$ might not be analytically tractable but $x^*$ values can be simulated (e.g., in the case of simulations with complex selection functions), we can evaluate
\begin{equation}
    \{x^*_{i,1}, \dots, x^*_{i,j}, \dots\} \sim \mathcal{S}_x(x_i, \sigma_i).
\end{equation}
Note that this relates to our original likelihood from above via
\begin{equation}
    P(x^*|x,\sigma) = \frac{P(x|x^*,\sigma)P(x^*)}{P(x)} \propto P(x|x^*,\sigma)P(x^*).
\end{equation}
Given a sample of $m$ simulated values, we can construct a Monte Carlo approximation of the integral as
\begin{equation}
    \int_{\Omega(x^*)} P_\phi(\theta, x^*) P(x | x^*, \sigma) \, {\rm d} x^* \approx \frac{1}{m} \sum_{j=1}^{m} P_\phi(\theta, x^*_j).
\end{equation}

The second challenge is missing data. One can also think of this as data where $\sigma_i \rightarrow \infty$, which means that we can assume $P(x|x^*,\sigma) \approx C$ over the entire domain $\Omega(x^*)$.
This gives
\begin{equation}
	P_\phi(\theta|x,\sigma=\infty) = P_\phi(\theta) \propto \int_{\Omega(x^*)} P_\phi(\theta, x^*) \, {\rm d} x^*.
\end{equation}

In practice, this integral is only done over some of the data. We can define this more explicitly by separating out $x = \{x_{\rm o}, x_{\rm m} \}$ and $\sigma = \{ \sigma_{\rm o}, \sigma_{\rm m}\}$ into observed $\{ x_{\rm o}, \sigma_{\rm o}\}$ and missing $\{x_{\rm m}, \sigma_{\rm m}=\infty\}$ values. Plugging in and combining/refactoring a few terms then gives
\begin{equation}
    P_\phi(\theta|x,\sigma) \propto \int_{\Omega(x^*)} P_\phi(\theta, x^*) P(x_{\rm o} | x^*_{\rm o}, \sigma_{\rm o}) \, {\rm d} x^*_{\rm o} {\rm d} x^*_{\rm m}.
\end{equation}

The strategy here will need to involve some approximations. We have already assumed that it is straightforward to simulate $x^*_{\rm o}$ values from $P(x_{\rm o} | x^*_{\rm o}, \sigma_{\rm o})$. If we can also simulate values from $P(x^*_{\rm m} | x^*_{\rm o})$, this implies that we can evaluate this integral using a Monte Carlo approach. More formally, if
\begin{equation}
  \begin{aligned}
    \{x^*_{{\rm o},j}\}_{j=1}^{m} & \overset{{\rm iid}}{\sim} P(x_{\rm o} | x^*_{\rm o}, \sigma_{\rm o}), \\
    x^*_{{\rm m},j} & \sim P(x^*_{\rm m} | x^*_{{\rm o}, j}),
\end{aligned}
\end{equation}
then our integral approximation becomes
\begin{equation}
    P_\phi(\theta|x,\sigma) \approx \frac{1}{m} \sum_{j=1}^{m} P_\phi(\theta | x^*_j) P(x^*_{{\rm o}, j}) = \frac{1}{m} \sum_{j=1}^{m} \frac{P_\phi(\theta, x^*_j)}{P(x^*_{{\rm m},j} | x^*_{{\rm o}, j})}.
\end{equation}

In terms of evaluating $P(x^*_{{\rm m},j} | x^*_{{\rm o}, j})$ to get our missing values, one strategy is to proxy this using a nearest neighbor search. Based on the neighbors, we can define a local density function $Q(x^*_{{\rm m},j} | x^*_{{\rm o}, j})$, and then simulate values of $x^*_{{\rm m},j}$ from that distribution. Assuming $P(x^*_{{\rm m},j} | x^*_{{\rm o}, j}) \approx Q(x^*_{{\rm m},j} | x^*_{{\rm o}, j})$, we finally get
\begin{equation}
P_\phi(\theta|x,\sigma) \approx \frac{1}{m} \sum_{j=1}^{m} \frac{P_\phi(\theta, x^*_j)}{Q(x^*_{{\rm m},j} | x^*_{{\rm o}, j})}.
\end{equation}

\end{document}